\documentstyle[12pt]{article}
\begin{document}
\newcommand{\ket}[1]{|#1>}
\newcommand{\tbox}
{\raise.3ex\hbox{$\widetilde{\lower.5ex\hbox{$\Box$}}$}}
\newcommand{\eqn}[1]{Eq. \ref{#1}}
\newcommand{\coho}{cohomology }
\newcommand{\ab}{antibracket }
\newcommand{\Psl}{\not\!\! P}
\newcommand{\half}{{\textstyle\frac{1}{2}}}
\newcommand{\cO}{{\cal O}}
\newcommand{\QED}{{\hspace*{\fill}\rule{2mm}{2mm}\linebreak}}
\newcommand{\dr}{\stackrel{\leftarrow}{\partial }\!\!{}}
\newcommand{\dl}{\stackrel{\rightarrow}{\partial }\!\!{}}
\renewcommand{\section}[1]{\addtocounter{section}{1}
\vspace{5mm} \par \noindent
  {\bf \thesection . #1}\setcounter{subsection}{0}
  \par
   \vspace{5mm} }
\newcommand{\sectionsub}[1]{\addtocounter{section}{1}
\vspace{5mm} \par \noindent
  {\bf \thesection . #1}\setcounter{subsection}{0}\par}
\renewcommand{\subsection}[1]{\addtocounter{subsection}{1}
\vspace{2.5mm}\par\noindent {\em \thesubsection . #1}\par
 \vspace{0.5mm} }
\renewcommand{\thebibliography}[1]{\section{References}
\list
 {\arabic{enumi}.}{\settowidth\labelwidth{[#1]}\leftmargin\labelwidth
 \advance\leftmargin\labelsep\addtolength{\topsep}{-4em}
 \usecounter{enumi}}
 \def\newblock{\hskip .11em plus .33em minus .07em}
 \sloppy\clubpenalty4000\widowpenalty4000
 \sfcode`\.=1000\relax \setlength{\itemsep}{-0.4em} }
\begin{titlepage}
\begin{flushright} preprint-KUL-TF-91/35
\end{flushright}
\vfill
\begin{center}
{\bf BATALIN-VILKOVISKY LAGRANGIAN QUANTISATION. $^1$}\\
\vspace{1.4cm}
{\bf Antoine VAN PROEYEN $^2$}\\
\vskip 0.3cm
Instituut voor Theoretische Fysica
\\Universiteit Leuven,
B-3001 Leuven
\end{center}
\vfill
\begin{center}
{\bf ABSTRACT}
\end{center}
\begin{quote}
The Lagrangian Batalin-Vilkovisky (BV) formalism gives the rules for
the quantisation of a general class of gauge theories which contain all the
theories known up to now. It does, however, not only give a recipe to
obtain a gauge fixed action, but also gives a nice understanding of the
mechanism behind gauge fixing. It moreover brings together a lot of
previous knowledge and recipes in one main concept~: the canonical
transformations. We explain the essentials of this formalism
and give related results on the superparticle.

Also anomalies (in general functions of fields and antifields) can be
obtained in this formalism, and it gives
the relation between anomalies in different gauges. A Pauli-Villars
scheme can be used to obtain a regularised definition of the
expressions at the one loop level.
The calculations become similar to those of Fujikawa
with the extra freedom of using arbitrary variables. A discrepancy between
anomalies in light-cone gauge of the Green-Schwarz superstring and in the
semi-light-cone gauge is discussed.
\vfill  \hrule width 5.cm
\vskip 2.mm
{\small
\noindent $^1$ to be published in the proceedings of "Strings and
Symmetries 1991", may 20-25, Stony Brook.\\
\noindent $^2$ Onderzoeksleider,
N.F.W.O. Belgium; Bitnet FGBDA19 at BLEKUL11}
\normalsize
\end{quote}
\begin{flushleft}
September, 1991
\end{flushleft}
\end{titlepage}
\addtocounter{section}{1}
\par \noindent
  {\bf \thesection . Introduction}
  \par
   \vspace{5mm}
Gauge theories are responsible for nearly all important elementary particle
theories
in the last 25 years. Quantisation of gauge theories is then also one of
the most important ingredients of our field of research. From the simple
gauge theory of QED to the types of gauge theories which we use nowadays
in string theories is a large development of quantisation of gauge
theories. A lot of new aspects came in, as there are soft algebras, open
algebras, reducible algebras, ... and various people found how to quantise
them using different types of ghosts, antighosts, ghosts for ghosts,
Nakanishi-Lautrup fields, Nielsen-Kallosh ghosts, ... .
The Lagrangian formalism of Batalin and Vilkovisky \cite{BV} summarises all
these developments in a few general concepts.

But the BV formalism does even more. By
rephrasing everything in terms of canonical transformations
\cite{BVcan,anombv}, the whole
picture of the quantisation procedure becomes more transparent. One can
understand how the physical variables from the classical theory are
represented by the cohomology of the BRST operator of the gauge fixed
theory. One can understand the meaning of gauge fixing in a new way, namely
as a canonical transformation from a classical 'extended' action to a
'gauge fixed' extended action.

Further also the quantum aspects of gauge theories can be treated in this
way, in particular the anomalies of the theory \cite{anombv}. It becomes
clear
what happens when one goes from one gauge to another, these being related
by a canonical transformation.

I will use during the talk the particle and the bosonic
string to explain the general ideas. These are of course very simple gauge
theories. But it is clear that the formalism can handle much more
complicated theories. That is what it was set up for.

I will first treat the classical (zero loop) aspects of the BV
formalism, and then discuss anomalies. For the first part, the
idea of presenting the BV formalism in this way was developed during a
collaboration with Eric Bergshoeff and Renata Kallosh on the superparticle
\cite{disp}. I will give some conclusions of that article at the end of
that part. The
necessary technical tools I learned from the Ph. D. thesis of Jean Fisch
\cite{Fisch},
which summarises some articles of himself and Marc Henneaux \cite{HPT}. The
part about
anomalies was first developed a few years ago in a collaboration with
Walter Troost and Peter van Nieuwenhuizen \cite{anombv}, and partly also in
collaborations including moreover Alvaro Diaz \cite{anomPV}, Machiko
Hatsuda \cite{measure}, and
Fiorenzo Bastianelli \cite{anomgs}. Recently we got new insights in a
collaboration with Taichiro Kugo, Stany Schrans and Walter Troost
\cite{anomas}. The parts in this text about anomalies have been written
down together with Walter Troost and Stany Schrans.
\section{Ingredients : antifields and antibrackets}
As announced, I will use as an example the bosonic relativistic particle in
$D$ dimensions.  The classical action is
given by:
\begin{equation}
S_{cl}(\phi^i) =\int dt\ P^\mu\dot {X}_\mu - {1\over 2}gP^\mu P_\mu
\end{equation}
The classical fields $\phi^i$ are the coordinates $X_\mu$ $(\mu = 1,
\dots ,D)$, their
conjugate momenta $P^\mu$ and the 1-dimensional metric
$g$. The dot indicates a time derivative.
The above action is invariant under general coordinate transformations
(with parameter $\xi$): $
\delta X^\mu = \xi P^\mu\ ,\ \delta g = \dot {\xi}$.

The physical variables are by definition the variables that remain on the
'stationary surface' (that is the surface in the
field space defined by the
field equations). However in gauge theories two solutions are said to be
equivalent when they are connected by
gauge transformations. \begin{equation}
\mbox{phys. obs.}=\frac{\mbox{soln. of field eqs.}}
{\mbox{gauge transf.}}.
\end{equation}

The problem for making quantum calculations in gauge theories is that the
propagator is not defined, or in other words, the
matrix\footnote{The notations
$\dl_A$ and $\dr_A$ indicate left and right derivatives with
respect to fields $\Phi ^A$. We have $\dr_A X=(-)^{A(X+1)}\dl_A X$,
where $(-)^A$ is $+$ if $\Phi ^A$ is a bosonic field, and is $-$ if $\Phi
^A$ is a fermionic field. When we
write no arrow it does not matter whether it is a right or left
derivative
(or it is a printing error). The index $A$ is a shorthand for all the
indices which the fields can have and for the space-time point. Sums over
$A$ will thus also involve integrations over space-time.}  $
S_{cl\ ij}\equiv \dl_i \dr_j S_{cl}$
is not invertible on the stationary surface. Its rank is $<n$ where $n$ is
the number of fields $\phi ^i$.
The path integral is then also not well defined. The aim
will be to replace this classical action by a new action which
\begin{enumerate}
\setlength{\itemsep}{-0.4em}
\item is gauged fixed : propagators can be defined.
\item has the same physical content.
\end{enumerate}
For the latter requirement we will define the physical variables
of the gauge-fixed action
as solutions of the field equations which are in
the cohomology of some BRST operator $\Omega $. The latter is a fermionic
operator which squares to zero modulo field equations (denoted by $\approx
$)~: $\Omega ^2 \approx 0$.
By the cohomology of this operator we mean the states $\ket{\Psi }$ which
satisfy \begin{equation}
\Omega \ket{\Psi }\approx 0   \label{BRSTcoho1}
\end{equation}
where 2 states are equivalent which differ on shell by a BRST exact state~:
\begin{equation}
\ket{\Psi }\sim \ket{\Psi '}\approx \ket{\Psi }+\Omega \ket{\chi }
\label{BRSTcoho2}\end{equation}

The essential ingredients in the BV procedure are antifields and
antibrackets. For all fields $\Phi ^A$ (including the classical fields
$\phi ^i$ introduced above, but this set of fields will be enlarged by
e.g.
ghosts), one introduces an antifield $\Phi ^*_A$. These have opposite
statistics as their corresponding fields.  In our example the fields
$\phi ^i$ are all bosonic, so the $\phi ^*_i$ are fermionic.
Further a ghost number is defined
such that for the classical fields $gh(\phi ^i) =0$, the (extended)
action
has ghost number zero, and for all fields $
gh(\Phi^*_A )= -gh(\Phi ^A)-1$.
The {\bf antibrackets} are defined
between two functions $F$ and $G$ of the fields and antifields by
\begin{equation}
(F,G) = \dr_A F\dl^A G - \dr^A F\dl_A G ,\label{abracket}
\end{equation}
where $\partial _A\equiv \frac{\partial }{\partial\phi ^A}$ and
$\partial ^A\equiv \frac{\partial }{\partial\phi ^*_A}$.
They satisfy graded commutation, distribution and Jacobi relations. For
these brackets fields and antifields behave as coordinates and momenta
\begin{equation}
(\Phi^A,\Phi^B)=0\ ;\hspace{0.5cm}
(\Phi^*_A,\Phi^*_B)=0;\hspace{0.5cm}
(\Phi^A,\Phi^*_B)=\delta^A_B.\label{eq:brackffs}
\end{equation}

Sometimes it is useful to write all fields and antifields in a general
notation $\{z^\alpha \}=\{\Phi ^A,\Phi ^*_A\}$ and define
\begin{equation}
\eta ^{\alpha \beta }\equiv (z^\alpha ,z^\beta ) \ ;\qquad
(F,G) = \dr_a F\eta ^{ab}\dl^b G .
\end{equation}

\sectionsub{Construction of the quantum action}
\subsection{Construction of the extended action}
We first extend the classical action to an extended action $S(\Phi ,\Phi
^*)$. It should satisfy
\begin{itemize}
\item $S_{cl}(\phi ) = S(\Phi ,0)$~: the classical
action is recovered when all antifields are set to zero.
\item The extended action should satisfy the {\bf master equation}, which
at the classical level reads $(S,S)=0$.
\item $S$ is a proper solution, which means that $S_{\alpha \beta }\equiv
\dl_\alpha \dr_\beta S$ is
a matrix of rank $N$ on the stationary surface where $N$ is the number of
fields $\Phi ^A$.
\end{itemize}
The last condition gives the possibility of a gauge fixed action. In fact,
the gauge fixed action will just consist of the same extended action, but
where we choose other coordinates as fields, such that the $N$ non-trivial
directions of $S^\alpha {}_\beta (z) \equiv \eta ^{\alpha
\gamma }S_{\gamma \beta }(z)$
are fields rather than antifields.

The second requirement implies that on the stationary surface $S^\alpha
{}_\beta (z)$ satisfies
\begin{equation}
S^\alpha {}_\beta (z)S^\beta {}_\gamma (z)\approx 0 \end{equation}
where $\approx $ means modulo field equations of the extended action. A
nilpotent matrix of size $2N\times 2N$ has rank  $\leq N$, and only rank
$N$ if all its zero modes are contained in the matrix itself. This implies
that we have for arbitrary smooth local functions
$v^\alpha (z)$ the implication \begin{equation}
S^\alpha {}_\beta (z)v^\beta (z)\approx 0\ \Rightarrow \ v^\beta (z)\approx
S^\beta {}_\gamma  w^\gamma (z) \end{equation}
for some local smooth functions $w^\gamma (z)$.

In our example we have thus to extend the action by terms depending on
antifields, because the gauge invariance implies that there is a zero
mode. We have to extend the action such
that this zero mode appears as $S^i{}_c$ where $c$ is a new field to be
introduced, which is the ghost of the symmetry. So the extended action is
\begin{equation}
S(\Phi,\Phi^*) =\int dt\ P^\mu\dot {X}_\mu - {1\over 2}gP^\mu P_\mu
+ X^*_\mu cP^\mu + g^* \dot c .\label{Sextpar}
\end{equation}
In this case we are already finished.
It is clear that the terms linear in
the antifields are the BRST transformations of the corresponding fields.
This extended action satisfies the master equation,
and it is proper. The master equation includes in this
case just the invariance of the action under the symmetry. But in more
complicated theories the same principles include also the closure of the
algebra and all the relations found in
the previous years about open algebras, and other similar complications.
The properness implies that all zero modes, ... have to be included. The
master equation can always be solved perturbatively in antifield number
(that is the same as the ghost number for the antifields and zero for
fields.)

When we have obtained the extended action, the physical variables are
represented by an \ab (AB) cohomology at ghost number zero.
\begin{equation}
\mbox{phys. variables}\ \Rightarrow \mbox{local AB coho at ghost nr. 0}.
\label{physABcoho}\end{equation}
This means the following. The operation which consists of taking an \ab
with $S$ is nilpotent. Indeed from the Jacobi identity
$(S,(S,F))=\half ((S,S),F)=0.$
The \ab \coho are the local functions $F$ which have ${\cal S}F=(F,S)=0$
and where two solutions are equivalent which differ by $(S,G)$, where
$G$ is any local function.
Cohomology is now defined with $=$ instead of $\approx $ in
eqs.(\ref{BRSTcoho1}-\ref{BRSTcoho2}). This will allow to do field
redefinitions which change the field equations.
For general gauge theories the fact that the physical variables are
equivalent to this \ab \coho is proven using the language of homological
perturbation theory in \cite{HPT,Fisch}.
\subsection{Canonical transformations}
In the example
it is easy to define the fields such that the action is gauge
fixed. We have at this point the bosons $X^\mu ,\ P^\mu ,\ g$ and $c^*$
and the fermions $X^*_\mu,\ P^*_\mu ,\ g^*$ and $c$. It is clear from
\eqn{Sextpar} that we have a non-trivial kinetic term between $g^*$ and
$c$. Thus we will define $g^*$ as a field, which we will call $b$~:
\begin{equation}
g^*=b\ ;\qquad g=1-b^*.          \label{cangb}
\end{equation}
This is a canonical transformation which means that the
transformation preserves the brackets~: calculating brackets in
the old or
new variables is the same, or in other words the new variables also
satisfy \eqn{eq:brackffs}. Therefore they also
conserve the master equation $(S,S)=0$.

Canonical transformations from $\{\Phi\Phi ^*\}$ to $\{\Phi'\Phi'^*\}$
for which the matrix $
\left. \frac{\dr\Phi^B}{\partial \Phi'^A}\right|_{\Phi'^*}$
is invertible,
can be obtained from a fermionic generating function $F(\Phi,\Phi'^*)$.
This useful formulation \cite{BVcan}
is explained in detail in the appendix of \cite{anombv}. The
transformations are defined by
\begin{equation} \Phi'^A=\frac{\partial F(\Phi , \Phi'^*)
}{\partial \Phi'^*_A}\hspace{2cm}
\Phi^*_A=\frac{\partial F(\Phi , \Phi'^*)}{\partial \Phi ^A}.
\label{Fcan}\end{equation}

Canonical transformations are an important part of the formalism.
In this one concept we find several steps which people do in quantisation
procedures.
\begin{itemize}
\item Point transformations are the easiest ones. This are just
redefinitions between the fields $\Phi '^A=f^A(\Phi )$. They are obtained
by $F=\Phi '^*_A f^A(\Phi )$ which thus determines the corresponding
transformations of the antifields. The latter replace the calculations of
the variations of the new variables.
\item Adding the BRST transformation of a function $s\Psi(\Phi )$ to the
action is obtained by a canonical transformation with $F=\Phi '^*_A\Phi ^A+
\Psi (\Phi )$. The latter gives
\begin{equation}
\Phi'^A = \Phi^A \ ;\qquad
\Phi ^*_A =\Phi'^*_A + \partial _A \Psi (\Phi ).
\label{gfermion}\end{equation}
\item Redefine the symmetries by adding equation of motions
('trivial symmetries'). This is obtained by
\begin{equation}
F=\Phi '^*_A\Phi ^A+ \Phi '^*_A\Phi '^*_B h^{AB}(\Phi )
\end{equation}
(the first term is the identity transformation).
\item Elimination of auxiliary fields can be done by canonical
transformations
(see appendix B of \cite{disp}). This procedure will then also give
the 'compensating transformations'. \end{itemize}
The canonical transformations keep by definition the master equation
invariant, and because they are non-singular, they also keep the properness
requirement on the extended action. Of course in the new variables, we do
not see the classical limit anymore. But the most important property is
that the antibracket cohomology is not changed. This should be obvious from
the definition.
\subsection{The gauge fixed theory}
After our transformation \eqn{cangb} we end up with
the following extended action:
\begin{equation}
S =\int dt\ P^\mu \dot {X_\mu} - \half P^2 + b\dot c + X^*_\mu cP^\mu
+ \half b^*P^2 . \label{Sgfpar}
\end{equation}
Now the action obtained by putting all antifields to zero is 'gauge fixed'.
For a gauge fixed action one can prove \cite{Fisch,HPT}
\begin{equation}
\mbox{local AB coho}\ \Rightarrow \ \mbox{local BRST
coho}.\label{ABcohoBRST} \end{equation}
The \ab \coho is now represented by the BRST \coho for some operator
\begin{equation}
\Omega \Phi ^A = \left.\frac{\dl S}{\partial \Phi ^*_A}\right|_{\Phi ^*=0}
\label{defOm}\end{equation}
The combination of \eqn{physABcoho} and \eqn{ABcohoBRST} imply that the
physical variables are represented by the local BRST cohomology at ghost
number zero.
In contradistinction to the antibracket cohomology we have to use the field
equations (of this gauge-fixed action) in the analysis of the BRST
cohomology. \subsection{Trivial variables}
We have done here the gauge fixing by just one canonical transformation~:
\eqn{cangb}. This is usually possible for actions which are linear in
derivatives. But in general it is not always possible to find suitable
covariant variables.
Therefore we add new trivial variables.

Trivial variables are a set of fields with corresponding
antifields, for which terms are added or subtracted from the extended
action.
They are separately solutions of the master equation and carry no
\ab cohomology. The simplest examples are adding a bosonic $\lambda $ and
fermionic $b$ and corresponding antifields by an extra term in $S$ of the
form $b^*\lambda $. One checks that this trivially satisfies the master
equation, that one has changed the number of fields + antifields by 4
and increased the rank of $S_{\alpha \beta }$ by 2, and that there is no
change in the \ab cohomology.
Other examples are just adding a bosonic $\lambda
$ with $S=\lambda ^2$ or a fermionic $b$ with $S=\left(b^*\right) ^2$.

The addition of such trivial $b,\ \lambda $ sectors is part of the scheme
which Batalin and Vilkovisky suggested \cite{BV} to obtain the gauge
fixed
action. Then they propose to do a canonical transformation of the type
\eqn{gfermion}. If a `{\it gauge fermion}' can be found which satisfies
certain conditions then the action is gauge fixed.
However, for a given gauge theory it is not clear
that such a gauge fermion exist. In other words, choosing the right
variables for the gauge fixed theory has an arbitrariness and there is no
guarantee that such a covariant basis exist.
\subsection{The superparticle}
As an example we mention that such a procedure has not yet been found for
the Brink-Schwarz (BS) superparticle action (for $D=10$) \cite{BS}
\begin{equation}
S_{cl} = P^\mu \dot X_\mu
-\theta\Psl \dot \theta -\half gP^{\mu}P_{\mu}.\label{BSact}
\end{equation}
The classical variables are the coordinates $X_\mu$, their conjugate
momenta $P^\mu$, the einbein $g$ and the fermionic variable
$\theta$ which is a ten-dimensional Majorana-Weyl spinor.
The action is infinitely reducible, which means that one needs an
infinite number of ghosts, called $\theta _{p0}$ with $p=1,...,\infty $
to be added to the classical variables in order to obtain the
minimal solution for the extended action \cite{Reninf,SBSP}.
Then one adds fields as suggested
by Batalin and Vilkovisky. That implies two infinite pyramids of fields.
Steps which people took to get a `gauge fixed action'
are now recognised as being no canonical transformations or
adding variables which are not `trivial' \cite{disp}.  Therefore the
BRST operator which was found for the  final gauge fixed action had extra
solutions for its cohomology which do not correspond to physical variables
of the classical action. The spinor variables in the gauge
fixed action are given in table~\ref{fieldssp} with their ghost number.
\begin{table}[htf]\caption{Spinor fields in the superparticle gauge-fixed
action.} \label{fieldssp}\begin{center}\begin{tabular}{ccccccc}\hline
$-3$&$-2$&$-1$            &0        &          1&2          &3  \\ \hline
    &    &                &$\theta $&           &           &   \\
    &    &$\bar\theta^{11}$&       &$\theta _{10}$ &           &   \\
    &$\bar\theta^{21}$&   &$\theta_{11}$&       &$\theta _{20}$&   \\
$\bar\theta^{31}$&&$\bar\theta^{22}$&&$\theta_{21}$&  &$\theta _{30}$\\
\vdots&\vdots&\vdots&\vdots&\vdots&\vdots&\vdots\\
\hline &    &                &$\bar\lambda ^{00}$&     &
& \\ &    &$\bar\lambda^{10}$&       &$\lambda_{11} $ &           &   \\
    &$\bar\lambda^{20}$&   &$\bar\lambda^{11}$&       &$\lambda_{21}$& \\
$\bar\lambda^{30}$&&$\bar\lambda^{21}$&&$\lambda_{22}$&
&$\lambda_{31}$  \\
\vdots&\vdots&\vdots&\vdots&\vdots&\vdots&\vdots\\
\end{tabular}\end{center}\end{table}
All the fields with 2 non-zero indices have been introduced as trivial
sectors. Nevertheless, there were counting arguments in favor of this set
of fields based on considerations about orthosymplectic symmetry
\cite{RenBars}. And indeed for this gauge fixed action another BRST
operator
has been found \cite{cohoart}, which has the `right \coho', that is the
fields occurring
in the $d=10$ super-Maxwell theory. However, this BRST operator does not
follow from a quantisation procedure on the BS action. The gauge fixed
extended action at this point is of the form
\begin{equation}
S_{gf}= \half\Phi ^A C_{AB}\dot \Phi ^B -\half P^2+ \Phi ^*_A C^{AB}\dl_B
\Omega  \label{SfromO}\end{equation}
where $C_{AB}$ is a constant non-singular matrix, with inverse $C^{AB}$ and
$\Omega $ is the BRST current. We can then perform again a canonical
transformation to a classical action. It is in fact the generalisation of
the inverse of \eqn{cangb}. All the fields of negative ghost number in
table~\ref{fieldssp} are replaced by their antifields. E.g. the antifields
of the fields of ghost number $-1$ become in this way fields of ghost
number 0, and will be called gauge fields (as there was $g$ in our example
of the particle). So after this canonical transformation which is the
inverse of a gauge fixing, all the fields of ghost number 0 are now
classical fields. The classical action is then \cite{neww}
\begin{eqnarray}
S_{cl} &=& P\dot X - \half gP^2 + \sum_{p=0}^\infty \lambda^p \dot
{\theta_p} \nonumber \\
&&-\sum_{p=0}^\infty \lambda^p \Psl\zeta^p - \sum_{p=0}^\infty
\{\lambda^{p+1} - \theta_{p+1}\Psl + \lambda^p
+\theta_p\Psl\}\eta _p \label{disp}
\end{eqnarray}
where $\lambda ^p\equiv \bar \lambda ^{p,p},\ \theta _p=\theta _{p,p},\
\zeta ^p=-\bar \lambda^*_{p+1,p}$ and $\eta _p=\bar \theta^*_{p+1,p+1}$.
The full extended action follows from straightforward application of the
quantisation principles. All the remaining fields in the table (or their
antifields for the negative ghost numbers) now occur as minimal ghosts or
ghosts for ghosts. There are a double infinite set of fields, of
symmetries, of zero modes of this symmetries, ..., which is the reason why
we have called this formulation of the superparticle the DISP (Doubly
Infinite Symmetric superParticle). It has the same physical variables as
the BS action, but allows a straightforward quantisation because in the
Hamiltonian language there are only first class constraints.
Other similar approaches, where the BS action has also been replaced by
another classical action, allowing a consistent quantisation,
have been given in \cite{ilketal}.
\section{Anomalies at the formal level}
In the full quantum theory, the master equation gets replaced by
\begin{equation}
(W,W)= 2i\hbar \Delta W,  \label{eq:qme}
\end{equation}
where \begin{equation}
\Delta\equiv (-)^A\dl_A\dl^A=\half (-)^\alpha \dl_\alpha \eta ^{\alpha
\beta }\dl_\beta .\end{equation}
$W$ is the quantum action which can be expanded in a loop expansion as
\begin{equation}
W=S+\hbar M_1 +\hbar ^2 M_2+\ldots .
\end{equation}
and we require that this action is local (see e.g. below for $M_1$).
The lowest order of \eqn{eq:qme} is the classical master equation.
At one-loop we have the equation
\begin{equation}
(M_1,S)=i\Delta S.         \label{masterh}
\end{equation}
The sum over $A$ in the definition of $\Delta $ involves an integral over
the space-time points, and thus for a local action $\Delta S$ is
proportional to $\delta (0)$. We will need therefore a regularisation
scheme to make sense of these expressions. That is what we should
expect in all quantum field theories. In the following section we will
talk about such a regularisation scheme, but first we will give here some
general results valid for a regularised definition of $\Delta $ such that
its fundamental properties related to the \ab algebra are preserved. We
will restrict ourselves here to one-loop.

While there is a proof that one can always find a solution to the classical
master equation $(S,S)=0$, there is no guarantee that there exist a local
solution for $M_1$ in \eqn{masterh}. Local means here that the
function $M_1$ is of the form
\begin{equation}
M_1= \int dx\ m\left(\Phi (x), \partial \Phi (x),\ldots \right)
\label{localM}\end{equation}
where there is a number $n$ such that there are no terms with more than $n$
derivatives on the fields. If there is such a solution, then there is no
problem with preserving the gauge symmetries in the quantisation at one
loop. If there
is no such solution, then we say that there is an anomaly. We define
\begin{equation}
{\cal A}(\Phi,\Phi ^*)=\Delta S +i(S,M_1)
\end{equation}
using any `local counterterm' $M_1$.
This `anomaly' has ghost number 1. In the usual cases it can be written as
$c^a{\it a}_a$ where $c^a$ are the ghosts. Then $<{\it a}_a>$ is the
anomaly in the corresponding symmetry in the sense that it gives the
change of the path integral under a change of the gauge fixing for that
symmetry. From this it is clear that by
choosing another local counterterm the anomalies can be moved to different
symmetries.

{}From the general properties of brackets and the $\Delta $ operation
it follows that, at least formally, \begin{equation}
({\cal A},S)=0.
\end{equation}
This is the consistency condition for the anomalies. We are thus
investigating `consistent anomalies'. On the other hand
\begin{equation}
{\cal A}\sim {\cal A}'={\cal A}+i(M,S)
\end{equation}
so that possible anomalies are solutions of the \ab \coho at ghost number
1. Remark that to classify the physical states as the \ab \coho at ghost
number zero, we were talking about local functions, while here we consider
integrals over space-time of local functions in the sense of \eqn{localM}.

As an example we can look at these equations for the bosonic string.  We
\cite{anomas} investigated what constraints are implied by the consistency
equations if we suppose that a candidate anomaly is of the form
\begin{equation}
\Delta S =\int d^2x \ d \left(g_{\alpha \beta },g^{* \alpha \beta },
c^\alpha , c^*_\alpha ,c,c^*\right)  \end{equation}
where $d$ is a functional of the fields indicated and a finite number of
derivatives of these fields. Here
$c^\alpha$ is the ghost for general coordinate transformations
and $c$ is the ghost for local dilatations.
By canonical transformations
one can first `trivialise' $c$ and the determinant of the metric. This
implies that up to counterterms all solutions of the consistency
equations can be written in terms of $c^\alpha $ and two remaining
components of the metric, for which we take
\begin{equation}
h_{++}= \frac{g_{++}}{g_{+-}+\sqrt{g}}
\end{equation}
and its $+\leftrightarrow -$ interchanged, and their antifields.
The only solutions
which can not be absorbed in a counterterm and which have the correct
dimensionality can be written as \begin{equation}
A_L\equiv \int d^2x \left(c^- +h_{++}c^+\right)  \partial ^3_- h_{++}
\end{equation}
and its partner $A_R$. If the right-left
symmetry is not broken, then we can only find an anomaly proportional to
$A_L +A_R$. Going back to the original variables $g_{\alpha \beta },\
c^\alpha $ and $c$, and by adding local counterterms $M$, this can be
written as \begin{equation}
A_L +A_R =-\half \int d^2x\ c\sqrt{g}R + (M,S).\label{consLRanom}
\end{equation}

This is the usual expression of the dilatation anomaly.  In the conformal
gauge $g_{\alpha \beta }=\eta _{\alpha \beta }$, this expression vanishes,
but this does not mean that there is no anomaly.  This can be avoided by
introducing a gauge choice $g_{\alpha \beta }=\rho _{\alpha \beta }$ where
$\rho $ is some background metric.  The anomaly is then a functional of
this $\rho $.  In the present scheme, this procedure does not make sense~:
the anomaly is a functional of the fields, not of background fields (and
can thus be shifted by local counterterms, which are also field-dependent).
The conformal gauge is obtained in a way analogous to \eqn{cangb} for the
particle~:
\begin{equation}
g^{*\alpha \beta }= -b^{\alpha \beta }\ ;\qquad g_{\alpha \beta
}= \eta _{\alpha \beta }+b^*_{\alpha \beta }.
\label{canstr}\end{equation}
The field $g$ occurring in \eqn{consLRanom} is then a function of the
antifields.  This re-interpretation allows one to determine the form of the
anomaly in any gauge, using canonical transformations.  So the next
question is~: how do these affect the expression for the anomaly~?

Canonical transformations do not leave the $\Delta $ operation invariant.
But we have
\begin{equation}
\Delta S-\Delta 'S=\half (S,\ln J)\ ;\qquad
J= sdet\ \frac{\partial(\Phi \Phi^*)}{\partial (\Phi' \Phi'^*)} .
\end{equation}
If $\ln J$ is a local expression then the anomaly does not change in the
cohomological sense. In the above case $J=1$, so there is no change at all
by going from the original to the gauge-fixed basis.
But changing to different variables with a non-local canonical
transformation, this formula gives formally the change in the anomaly.
\sectionsub{Regularisation}
\subsection{Introducing and eliminating Pauli-Villars fields}
We will use a Pauli-Villars (PV) regularisation. It will allow us to make
contact with the work of Fujikawa \cite{Fujikawa} on obtaining anomalies
from the non-invariance of the measure.
The consistency of the PV scheme will then imply that the obtained
expressions satisfy the consistency equations, a point which is very
unclear in Fujikawa's method.

To start, we introduce PV partners for all fields and antifields. So we
have \begin{equation}
z^\alpha = \{\ \ \Phi ^A\ ,\ \ \Phi ^*_A\ \}\ ;\qquad
w^\alpha = \{\ \ \chi ^A\ ,\ \ \chi ^*_A\ \}
\end{equation}
where the latter are the PV fields.
For the calculations at one loop it
is sufficient to give the extra recipe that loops of PV fields produce an
extra minus sign which reflects itself in a modified definition of $\Delta
$~:
\begin{equation}
\Delta \equiv (-)^A \dl_A \dl^A -(-)^A \frac{\dl}{\partial \chi^A}
\frac{\dl}{\partial \chi_{A}^*}.   \label{DelPV}
\end{equation}
To obtain this sign one can modify the definition of the path integral over
these fields \cite{anomPV}, or else introduce extra sets of fermions and
bosons. The first method is certainly the simplest when we are
only looking to one loop. The PV fields then have the same statistics
as the ordinary fields, but one can say that the integration over
these fields in the path integral is
defined differently, such as to produce a minus sign. This is
consistent for fields which occur only in loops, i.e. quadratically
in the action.  The second method produces the minus in the loop by having
opposite statistics for PV fields.  E.g.  for regularising a real boson
field, we have to
introduce a fermion. But as the kinetic operator of a boson is
a symmetric operator, this would vanish for a fermion. Therefore one
has to introduce a complex fermion (or 2 reals). Then one has
over-compensated the loop of the original boson in the regularisation
procedure. One thus introduces an extra PV boson. So in summary, one has
introduced
e.g. for a boson field $\phi $, 3 PV fields~: the fermions
$\chi _1$ and $\chi _2$ and the boson $\chi _0$. Each of them have also
their antifields. But again, this complication can be forgotten when
considering 1 loop anomalies~: one may just treat the PV fields as
having the same statistics, and insert the sign by hand.

The PV fields are introduced as trivial systems in the limit $M\rightarrow
\infty $. We add to the action a term $M^2 \chi ^AT_{AB}\chi ^B$ where $T$ is
an invertible matrix. This implies that the $\chi ^*$ fields are not
invariant under $\cal S$ while the $\chi $ fields are in the image of this
operator. In the language of Feynman graphs, these fields are thus very
massive. Also, these fields should provide a regularisation of the original
action.  Therefore, neglecting the mass terms, they should produce the same
vertices as the original action.
The regularisation now consists in postponing to take the limit
$M\rightarrow \infty $ to the end of the calculation. We will define
\begin{equation}
S^{PV}=S^{PV}_0+S^{PV}_M= \half w^\alpha S_{\alpha \beta }w^\beta -\half M^2
\chi ^A T_{AB}\chi ^B  \label{SPV}\end{equation}
Let us first look at the massless part.
It is the $w^2$ part of the
action \begin{equation}  S^{reg}_0=
\half \left( S(z +w ) +S(z -w )\right)=S(z)+S^{PV}_0+ {\cal O}(w^4)
\end{equation}
which automatically satisfies then the classical master equation. When
looking
only at one loop (the only case which we consider here), one can forget all
terms of order $w^4$. In the formulation with 3 PV fields for each
ordinary field, of which $\chi _1^A$ and $\chi _2^A$ have opposite
statistics from $\phi ^A$ and $\chi ^0$ has the same statistics, we
would write
\begin{equation}
S^{PV}_0 -\dl_1 \dr_2 S +\half\dl_0 \dr_0 S.
\end{equation}
where we defined the operators
\begin{equation}
\partial _0 =\chi _0^A\partial _A +\chi ^*_{0A}\partial ^A\ ;\
\partial _1 =\chi _1^A\partial _A +\chi ^*_{2A}\partial ^A\ ;\
\partial _2 =\chi _2^A\partial _A +\chi ^*_{1A}\partial ^A .
\end{equation}
The notation $\dr_2$ implies also that the corresponding $w$ fields
appear at the right of $S$.
This satisfies the master equation, and also $\Delta S^{PV}=0$, where we do
not have to modify $\Delta $ this time.

We can introduce this action from the start, even
before gauge fixing. The reason is that
introducing a PV regulator in this way 'commutes' with a canonical
transformation. By this we mean the following. Suppose that one has
introduced the PV fields using
the above prescription and then performs a canonical transformation. Any
canonical transformation on the $z$ fields can be generalised to a
canonical transformation on the $z$ and $w$ fields such that after this
canonical transformation the same result is obtained as when introducing
the PV sector only after the canonical transformation.
So we have the following scheme
\begin{equation}
\begin{array}{ccc}
S(z)&\stackrel{\mbox{can. transf.}}
{\longrightarrow}&\tilde S(\tilde z)
\\ \downarrow PV && \downarrow PV \\
S(z)+\half w^\alpha S_{\alpha \beta }w^\beta
&\stackrel{\mbox{can. transf.}}
{\longrightarrow}&\tilde S( z')
+\half w'^\alpha \tilde S_{\alpha \beta} w'^\beta
\end{array}
\end{equation}
So this part is fixed immediately when giving the theory.

On the other hand, for the mass term there is a lot of arbitrariness.
First of all we have separated here fields from antifields.
This should be done suitably for gauge fixing,
i.e. $S_{AB}$ should have rank $N$. But once
this has been specified $T$ is an arbitrary non-singular matrix,
which may
depend on fields or antifields. We will claim that the results for the
anomalies for different choices of $T$ ('different regularisations')
differ only by the variations of local counterterms. In other words,
the anomalies do not change in the cohomological sense.

Let us now look again at the full master equation for the regularised
action
\begin{equation}
S^{reg}=S^{reg}_0 +S^{PV}_M = S(z)+S^{PV}+ {\cal O}(w^4)
\end{equation}
First of all, due to the definition \eqn{DelPV}
one has $\Delta S={\cal O}(w^2)$. But the anomalies now come from the
violation of the `classical master equation' due to the mass terms. This
will be proportional to $w^2$. Removing the PV fields by integrating them
out will replace $w^2$ by a term of order $\hbar $.

The violation of the master equation (to order $w^2$) is given by
\begin{eqnarray}
-i\hbar { \cal A}&=&\half (S^{reg},S^{reg})-i\hbar
\Delta S^{reg}\nonumber\\ &=&(S^{PV}_M,S^{reg}_0)+ {\cal O}(w^4)
+\hbar{\cal O}(w^2) \nonumber\\
&=&-M^2\chi ^AT_{AB}S^B{}_\alpha w^\alpha -\half M^2\chi ^A
\left(T_{AB},S(z)\right) \chi ^B (-)^B\nonumber\\ &&
+{\cal O}(w^4) +\hbar{\cal O}(w^2),
\end{eqnarray}
where $S^B{}_\alpha$ is the left derivative of $S$ w.r.t. $\Phi ^*_B$ and
right w.r.t. $z^\alpha $. Now we remove the PV fields. In a path integral this
would mean that the $\chi \chi $ terms are replaced by their propagator
(which gives a $\hbar $). The $\chi ^*$ terms are dropped in this step. We
follow this idea and defining
\begin{equation}
{\cal O}^A{}_B\equiv T^{-1\ AC}S_{CB} \ ; \qquad K^A{}_B\equiv S^A{}_B ,
\label{defOK}
\end{equation}
we obtain
\begin{equation}
{\cal A}=\left(K+\half T^{-1}(T,S)
(-)^B\right)^A_{\ B}\left(\frac{M^2}
{M^2-\cO}\right)^B_{\ A}(-)^A.\label{eq:kwado}
\end{equation}
Note that for gauge theories with a closed algebra $K$
is the matrix of the derivatives of the transformation of fields w.r.t.
the fields. In the limit $M\rightarrow \infty$ we take the trace of this
expression (for a field independent matrix $T$).
Now we will regularise this using the regulator $\cal O$.

The PV fields were a way to obtain a regularised definition of $\Delta S$.
Without the PV fields $(S,S)=0$ and $\Delta S\neq 0$, and we define
$\Delta S$ as the expression in \eqn{eq:kwado}. Note that if we first
take the limit $M\rightarrow \infty $ (and for $T$ a constant matrix),
it corresponds indeed to the unregularised definition of $\Delta S$.

\subsection{The integrals and the Fujikawa regularisation}
We then replace the propagator by an exponential function. This will allow
us to make contact with the Fujikawa calculation of anomalies
\cite{Fujikawa}. \begin{equation}
\frac{1}{1-{\cal
O}/M ^2} = \int_0^\infty d \lambda\ \exp \left( \lambda {\cal
O}/M ^2 \right) \exp (-\lambda) . \label{eq:lapl} \end{equation}
For a local action, we split $A=(a,x),\ B=(b,y)$ and
\begin{equation}
(K +\half  T^{-1}(T,S)(-)^B)^A_{\ B}=J^a{}_b(x)\delta (x-y) =
J^{\dagger a}{}_b(y)\delta (x-y) \label{jacobian}\end{equation}
where $J$ is some differential operator.
Also for the propagator we have \begin{equation}
\cO^A{}_B={\cal R}^a{}_b(x) \delta (x-y) . \label{defcalR}
\end{equation}
The anomalies are at this point
\begin{eqnarray}
{\cal A}&=&\int_0^\infty d\lambda\ e^{-\lambda}\int dx\int dy \ str\
J^\dagger(y)\delta (x-y)\cdot \exp \left(\frac{ \lambda {\cal
R}(y)}{ M^2} \right)\delta (x-y)\nonumber\\
&=&\int_0^\infty d\lambda\ e^{-\lambda}\int dx\int dy \ str\
\delta (x-y)J(x) \exp \left(\frac{ \lambda {\cal
R}(x)}{ M^2} \right)\delta (x-y) \label{anomdeldel}
\end{eqnarray}
where $J$ and $\cal R$ are now considered as matrices in the $a,b$, and
$str$ denotes a supertrace over these indices. The $\cdot$ indicates that
derivatives in the operators do not act further.
Several lemma's have been obtained to calculate expressions of the
form \eqn{anomdeldel}. I
believe that the one in \cite{anomgs} contains most of the others.
However, the formulas in \cite{Gilkey} do in fact contain all this,
and could be used to generalise them even further.
One obtains an expansion in $M^2$. The divergent terms when
$M^2\rightarrow\infty$ can usually be eliminated by a local counterterm
$M_1$, but the PV procedure introduces several copies of PV fields with
masses adapted such that these terms disappear anyway. So we can forget
about them. One discards also the terms which vanish in this limit, and
the result is thus independent of $M^2$. From the last expression, one
can see that the $M^2$ independent terms have only $\lambda $ occurring
in the first factor, and the integral over $\lambda $ gives thus 1.

As already mentioned, the above regularisation scheme is very similar to
Fujikawa regularisation
\cite{Fujikawa} and to the heat kernel approach \cite{heatk}. These
approaches have to
 use {\bf special variables} in order to avoid anomalies in `preferred
symmetries'. The Jacobian which is
regularised in the Fujikawa approach corresponds to the first term in
\eqn{jacobian}. By including the second term in that equation we can avoid
this restriction to the `Fujikawa variables'.

\subsection{The Green-Schwarz superstring and the light-cone gauge}
A contradiction with the above statements on gauge-independence of
anomalies seems to exist in the Green-Schwarz superstring.  The classical
action can be gauge-fixed in the light-cone gauge \cite{GSlc} or in the
so-called semi-light-cone gauge \cite{GSslc}, both of which destroy the
manifest rigid space-time super-Poincar\'e invariance.  In the former there
are no space-time anomalies.  In the latter the local fermionic
$\kappa$-symmetry is fixed by the unitary gauge $\Gamma^+\theta =0$,
while the other local worldsheet symmetries are covariantly quantized.
It was claimed by Kraemmer and Rebhan that anomalies
in the semi-light cone approach do not cancel \cite{rebhan}.
On the other hand, M.  Chu \cite{chu} claimed that in a Hamiltonian
formulation, the definition of the Lorentz generators can be changed in
order $\hbar $ such that anomalies disappear.
Using our methods (although we did not yet completely formulate it in the
way described above) we redid the calculation \cite{anomgs} and obtained
${\cal A}=-12 \frac{1}{24\pi }A_L$.
Adding local counterterms this anomaly could still be moved to an
anomaly in global Lorentz symmetries, but it could not be canceled.

There seems therefore a contradiction between gauge independence and the
different results in light-cone and semi-light-cone gauge. The difference
between these two gauges is in the bosonic sector. The arguments why they
should be the same should go by using canonical transformation as in the
general theory written above. One could
even ask the question just for the bosonic string. The
calculation of anomalies in the conformal gauge,
which was reviewed above gives as
a final result that anomalies cancel for $D=26$. On the other hand in the
light cone gauge this requirement comes from the analysis of the quantum
rigid space-time Lorentz algebra. It only closes when $D=26$. Are these
two calculations connected~?
Is there a canonical transformation between both~?
We do not have a final answer to this question, but it seems that one can
not avoid non-local canonical transformations to eliminate the ghosts in
going to the usual light-cone gauge fixed action \cite{anomas}.
Therefore, we can see no contradiction with the statements above.
If one does not perform local canonical transformations, it is not
clear that anomalies should be conserved when going to the light-cone
gauge.

\section{Conclusions}
The Batalin-Vilkovisky formalism
is a convenient framework for the quantisation of gauge theories.
One first builds an extended action, function of fields and
antifields which satisfies 3 requirements~: a classical limit, the master
equation and a properness condition. The latter implies that there are
enough non-trivial directions in the Hessian of this extended action,
such that when choosing (by a canonical transformation) the `fields' as
those directions in the extended space and antifields as the other
directions, one obtains a gauge fixed action. Using antibracket cohomology
one can prove that this action describes the same physical states.

For the quantum theory one needs another master equation. A theory has no
anomalies if a local action can be found which satisfies this equation.
Possible consistent anomalies are solutions of antibracket cohomology
equations at ghost number 1. For
local actions, the definition of the operation $\Delta $ which occurs in
the master equation needs regularisation. By introducing Pauli-Villars
fields one can motivate the regularised definition
\begin{equation}
\Delta S=\lim_{M\rightarrow \infty } \int dx\int dy \ str\
\delta (x-y)J(x) \exp \left(\frac{ \lambda {\cal
R}(x)}{ M^2} \right)\delta (x-y)
\end{equation}
where infinite terms are removed.
$J$ and $\cal R$ are given by Eq.\ref{defOK},\ref{jacobian},
\ref{defcalR} using an
arbitrary matrix $T$ which `determines the regularisation'.
The anomaly is then a function of fields
and antifields, and does not change under local canonical transformations
(in the cohomological sense), in particular they are gauge independent.
They are also independent of the regularisation,
in particular of the choice of $T$.

For the superparticle and for the superstring we found difficulties with
the Brink-Schwarz and the Green-Schwarz actions. For the BS superparticle,
which is a good formulation in the light-cone gauge, we
could not find canonical transformations or the right trivial variables to
obtain a gauge fixed action. We found however another action, the DISP
superparticle, which describes at the classical level the same physical
states, and allows a straightforward quantisation \cite{disp}. Other
modified suitable actions have been found by other groups
\cite{ilketal}. For the superstring also the Green-Schwarz
action, which is a good formulation in the light-cone gauge, has anomalies
in a semi-covariant gauge (covariant in the bosonic sector).
This is not in contradiction with the general results, as going from the
covariant to the light-cone gauge involves
non-local canonical transformations. So far also
no gauge-fixing procedure has been found for the Green-Schwarz action.
Probably we also need in this case a modified action similar to those for
the superparticle.

\end{document}